\documentclass[aps,prl,twocolumn,tightenlines,superscriptaddress,showpacs]{revtex4-1}
\usepackage{epsfig,rotating,amsmath}
\usepackage{multirow}

\begin{document}

\title{Constraints on Skyrme Equations of State from Properties
of Doubly Magic Nuclei and Ab-Initio Calculations of Low-Density
Neutron Matter}

\author{B. Alex Brown}
\affiliation{National Superconducting Cyclotron Laboratory and Department
of Physics and Astronomy, Michigan State University, East Lansing,
Michigan 48824-1321, USA}

\author{A. Schwenk}
\affiliation{ExtreMe Matter Institute EMMI,
GSI Helmholtzzentrum f\"ur Schwerionenforschung GmbH, 64291 Darmstadt, Germany}
\affiliation{Institut f\"ur Kernphysik, Technische Universit\"at Darmstadt,
64289 Darmstadt, Germany}

\begin{abstract}
We use properties of doubly-magic nuclei and ab-initio calculations of
low-density neutron matter to constrain Skyrme equations of state
for neutron-rich conditions. All of these properties are
consistent with a Skyrme functional form and a neutron-matter
equation of state that depends on three parameters. With a reasonable
range for the neutron-matter effective mass, the values of the two other
Skyrme parameters are well constrained. This leads to predictions for
other quantities. The neutron skins
for $^{208}$Pb and $^{48}$Ca are predicted to be 0.182(10) fm and
0.173(5) fm, respectively. Other results including the
dipole polarizability are discussed.
\end{abstract}


\maketitle

The properties of the neutron equation of state (EOS)
are important for understanding
neutron skins and neutron stars \cite{skin}, \cite{steiner13}, \cite{hebeler13}.
Recently an extensive study was made of the constraints on Skyrme
energy-density functionals (EDFs) provided by the properties of
nuclear matter \cite{dutra}.
The standard form and the parameters of the Skyrme functional
are given in \cite{dutra}. Out of several hundred Skyrme EDFs, the
16 given in Table VI of \cite{dutra} called the CSkP set best
reproduced a selected set of empirical nuclear matter properties.
Five of these were eliminated \cite{dutra} since they gave transitions
to spin-ordered matter around densities of $  \rho = 0.25  $ fm$^{-3}$. One of the
remainder (LNS) was unstable for finite nuclei. The remaining 10 are
those given in Table~I and labeled with their name and order in
Table~VI of \cite{dutra}. To this list we add
the commonly used SLy4 \cite{sly4} and SkM* \cite{skms} functionals.
These 12 EDFs cover a reasonable range of values for the
symmetric-nuclear-matter
effective mass ($  m^{*}/m  $ = 0.70-1.00) and
incompressibility ($  K_{m}  $ = 212-242 MeV) as compared to
values extracted from the energy of the giant monopole
resonances ($  K_{m}  $ = 217-230 MeV) \cite{sag}.

In \cite{bab} these 12 EDFs were refined by a fit to properties
of the doubly magic nuclei
$^{16}$O, $^{24}$O, $^{34}$Si $^{40}$Ca, $^{48}$Ca, $^{48}$Ni,
$^{68}$Ni, $^{88}$Sr, $^{100}$Sn, $^{132}$Sn, and $^{208}$Pb.  The
properties included binding energies, single-particle energies \cite{skx},
root-mean-square (rms)
charge radii and rms neutron radii. The most experimentally uncertain
of these are the rms neutron radii. With a fixed value for the neutron
rms radius of $^{208}$Pb, the
$  t_{0}, t_{1}, t_{2}, t_{3}, x_{0}  $, $  x_{3}  $ and $  W  $ (the spin-orbit strength)
Skyrme parameters are well determined.

When the neutron skin of $^{208}$Pb is allowed to vary over the
range of $  R_{np} = \sqrt{<r_{n}^{2}>} - \sqrt{<r_{p}^{2}>} = 0.16  $ to
0.24 fm, all of the EDFs cross at a value of the neutron density of
about 0.10 fm$^{-3}$ with a value of $  E/N = 11.3(8)  $ MeV \cite{bab}. The slope
at this point depends on the neutron skin \cite{skin}, \cite{bab}.
The resulting neutron EOS are shown here in Fig.~1. They are
compared with the upper and lower values of the range from
next-to-next-to-next-to-leading order (N$^{3}$LO) calculations of
neutron matter \cite{N3LO} (dashed lines). We observe that the Skyrme
EOS obtained with a neutron skin of 0.20 fm are in best agreement
with the theoretical N$^{3}$LO range.

\begin{figure}
\includegraphics[scale=0.5]{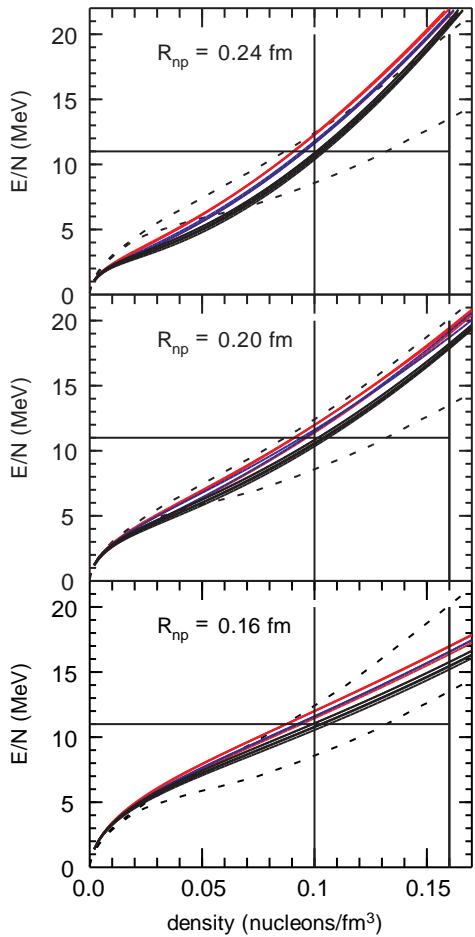}
\caption{The neutron EOS with $  m^{*}_{n}/m = 0.90  $ and with $  R_{np}  $
fixed at 0.16, 0.20 and 0.24 fm compared to the N$^{3}$LO neutron
matter band from \cite{N3LO} (dashed lines). The black lines are those
with symmetry nuclear matter values of
$  m^{*}/m \approx 1.0  $ and the red lines with
$  m^{*}/m = 0.70-0.85  $. The blue lines are for SLy4 and SkM*.}
\end{figure}

\begin{figure}
\includegraphics[scale=0.5]{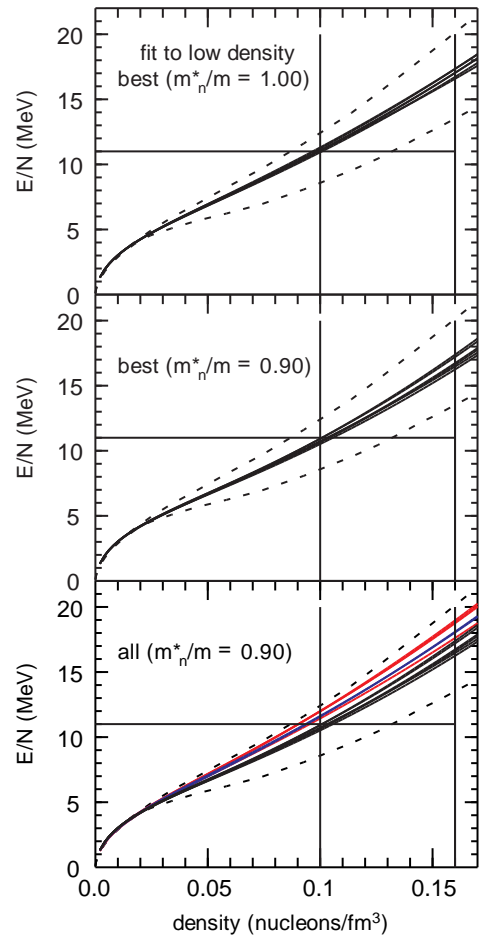}
\caption{The neutron EOS from Skyrme fits that include points from
ab-initio calculations of low-density neutron matter, compared to the
N$^{3}$LO neutron matter band from \cite{N3LO} (dashed lines).
The black lines are those
with symmetry nuclear matter values of
$  m^{*}/m \approx 1.0  $ and the red lines with
$  m^{*}/m = 0.70-0.85  $. The blue lines are for SLy4 and SkM*.}
\end{figure}

In this Letter, we follow up on this observation and investigate as
additional constraints the results from ab-initio calculations of
neutron matter. We simultaneously fit the experimental properties of
doubly-magic nuclei and the theoretical properties of low-density
neutron matter. We do not include any data on the neutron radii of
nuclei. Rather as a result of the simultaneous fit we will be able
make predictions for the neutron EOS and properties associated with it
such as the neutron radii.

We include theoretical data points for the neutron-matter energy at
three densities $  \rho_{n}=0.01, 0.02  $, and $  0.04  $ fm$^{-3}$. The
range for the energy per particle is based on the recent N$^{3}$LO
calculations including two-, three- and four-nucleon interactions in
chiral effective field theory (EFT) \cite{N3LO}, where the N$^{2}$LO
energies and the results with renormalization-group-evolved
interactions \cite{heb}, \cite{hebeler13} are also within this range. In
addition, we include in the energy range the results of first Quantum
Monte Carlo (QMC) calculations with local chiral EFT interactions at
NLO and N$^{2}$LO \cite{QMCchiral} (without three-nucleon forces,
which are small at low densities). At the lowest density point,
$  \rho_{n}=0.01  $ fm$^{-3}$, the energy range is
$$
[E/N](\rho_{n}=0.010 \, {\rm fm}^{-3}) = 3.00(13) \, {\rm MeV},       \eqno({1})
$$
which overlaps with the results from NLO lattice simulations that
yield around 3.1 MeV \cite{NLOlattice}. At the second point,
$  \rho_{n}=0.02  $ fm$^{-3}$, the N$^{3}$LO range is
$  E/N  $=4.14-4.34 MeV and the chiral QMC range is
4.39-4.60 MeV. These overlap with the variational calculations of
Akmal {\it et al.} based on the Argonne $  v_{18}  $ NN and the UIX 3N
potentials that give 4.35-4.45 MeV, where the range is due to
including boost corrections \cite{Akmal}. In addition, we consider the
Auxiliary Field Diffusion Monte Carlo (AFDMC) results based on the
Argonne $  v'_{8}  $ NN and the same UIX 3N
forces \cite{Gandolfi}. Extrapolating these from their lowest density
result at $  \rho_{n}=0.04  $ fm$^{-3}$ using their fit gives around
4.6 MeV. Therefore, we adopt as a conservative range for the energy
per particle at $  \rho_{n}=0.02  $ fm$^{-3}$
$$
[E/N](\rho_{n}=0.020 \, {\rm fm}^{-3}) = 4.37(23) \, {\rm MeV} \,.       \eqno({2})
$$
At the density $  \rho_{n}=0.04  $ fm$^{-3}$, the combined chiral EFT range
is 5.40-6.65 MeV, which overlaps with the results of Akmal {\it et al.}
that give 6.23-6.45 MeV, whereas the AFDMC results are $  6.79(1)  $ MeV.
Therefore, we take
$$
[E/N](\rho_{n}=0.030 \, {\rm fm}^{-3}) = 6.1(7) \, {\rm MeV} \,.       \eqno({3})
$$

\begin{table*}
\begin{center}
\caption{Properties of the fitted Skyrme functionals. The effective
mass $  m^{*}_{n}/m  $ in neutron matter at $  \rho_{n}=0.10  $ fm$^{-3}$ is
constrained to be $  0.9  $ in the first part of the table and $  1.0  $ in the
second part. The symmetry energy $  J  $, its density
derivative $  L  $, the symmetry-energy incompressibility
$  K_{s}  $, the symmetric-nuclear-matter incompressibility
$  K_{m}  $ and effective mass  $  m^{*}/m  $ are evaluated at
$  \rho=0.16  $ fm$^{-3}$. The mean value is from the entire set, and the
``best" value (b) is for the six cases that give the best fit to the
data.}
\begin{tabular}{|c|c||c|c|l|c|c||c|c|c||c|c|c||c|c|}
\hline
name & & $  \sigma  $ & $  m_{n}^{*}/m  $ & $  \chi^{2}  $ & $  K_{m}  $ & $  m^{*}/m  $ & $  a_{n} 
 $
& $  b_{n}  $ & $  d_{n}  $ & $  J  $ & $  L  $ & $  K_{s}  $ & $  R_{np}  $ & $  R_{np}  $ \\
& &  & & & (MeV) & & (MeV & (MeV & (MeV & (MeV) & (MeV) & (MeV) & (fm) & (fm) \\
& & & & &  & & fm$^{3}$) & fm$^{3 \gamma }$) & fm$^{5}$) &  & & & $^{208}$Pb & $^{48}$Ca \\
\hline
\hline
KDE0v1 & s3 & 1/6 & 0.90 & 1.81& 216 & 0.79 & $  -455  $ & 422 & 145 & 34.9 & 61 & $  -130  $ & 
0.192 & 0.172 \\
NRAPR & s6 & 0.14 & 0.90 & 2.60 & 225 & 0.85 & $  -534  $ & 509 & 133 & 35.1 & 61 & $  -142  $ & 
0.193 & 0.178 \\
Ska25 & s7 & 0.25 & 0.90 & 0.91 (b) & 219 & 0.99 & $  -360  $ & 328 & 122 & 32.5 & 51 & $  -138  $ 
& 0.176 & 0.170 \\
Ska35 & s8 & 0.35 & 0.90 & 0.80 (b) & 244 & 1.00 & $  -317  $ & 315 & 123 & 32.8 & 54 & $  -144  $ 
& 0.180 & 0.172 \\
SKRA & s9 & 0.14 & 0.90 & 1.64& 212 & 0.79 & $  -504  $ & 463 & 138 & 33.7 & 55 & $  -139  $ & 
0.181 & 0.172 \\
SkT1 & s10 & 1/3 & 0.90 & 0.84 (b) & 242 & 0.97 & $  -324  $ & 326 & 124 & 33.3 & 56 & $  -140  $ & 
0.183 & 0.172 \\
SkT2 & s11 & 1/3 & 0.90 & 0.86 (b) & 242 & 0.97 & $  -331  $ & 338 & 125 & 33.5 & 58 & $  -135  $ & 
0.186 & 0.174 \\
SkT3 & s12 & 1/3 & 0.90 & 0.80 (b) & 241 & 0.98 & $  -322  $ & 314 & 134 & 32.7 & 53 & $  -144  $ & 
0.179 & 0.172 \\
SQMC750 & s15 & 1/6 & 0.90 & 2.41& 228 & 0.71 & $  -467  $ & 447 & 123 & 34.8 & 59 & $  -148  $ & 
0.190 & 0.176 \\
SV-sym32 & s16 & 0.30 & 0.90 & 0.86 (b) & 237 & 0.91 & $  -335  $ & 313 & 123 & 32.3 & 51 & $  -148 
 $ & 0.176 & 0.174 \\
\hline
SLy4 & s17 & 1/6 & 0.90 & 1.97& 224 & 0.70 & $  -450  $ & 412 & 136 & 34.1 & 56 & $  -145  $ & 
0.184 & 0.174 \\
SkM* & s18 & 1/6 & 0.90 & 1.69& 218 & 0.78 & $  -473  $ & 450 & 120 & 34.2 & 58 & $  -139  $ & 
0.187 & 0.175 \\
\hline
mean & & & & & & & & & & 33.8(13) & 56(5) & $  -138(8)  $ & 0.184(9) & 0.174(4) \\
mean (b) & & & & & & & & & & 32.9(4) & 54(4) & $  -141(4)  $ & 0.180(4) & 0.172(2) \\
\hline
\hline
Ska25 & s7 & 0.25 & 1.00 & 0.85 (b) & 218 & 0.99 & $  -386  $ & 424 & 2 & 32.6 & 48 & $  -165  $ & 
0.173 & 0.168 \\
Ska35 & s8 & 0.35 & 1.00 & 0.73 (b) & 244 & 1.00 & $  -333  $ & 419 & $  -3  $ & 33.1 & 53 & $  
-165  $ & 0.179 & 0.171 \\
SkT1 & s10 & 1/3 & 0.99 & 0.76 (b) & 241 & 0.97 & $  -341  $ & 423 & 0 & 33.4 & 53 & $  -163  $ & 
0.181 & 0.170 \\
SkT2 & s11 & 1/3 & 1.00 & 0.76 (b) & 230 & 0.97 & $  -338  $ & 413 & 4 & 33.1 & 52 & $  -164  $ & 
0.179 & 0.170 \\
SkT3 & s12 & 1/3 & 1.00 & 0.73 (b) & 241 & 0.98 & $  -337  $ & 408 & 3 & 32.8 & 50 & $  -166  $ & 
0.176 & 0.170 \\
SV-sym32 & s16 & 0.30 & 1.02 & 1.04 (b) & 242 & 0.91 & $  -364  $ & 450 & $  -21  $ & 33.4 & 51 & $ 
 -176  $ & 0.178 & 0.173 \\
\hline
mean (b) & & & & & & & & & & 33.2(2) & 50(3) & $  -170(7)  $ & 0.178(4) & 0.171(3) \\
\hline
\hline
overall & & & & & & & & & & 33.1(20) & 52(9) & $  -180(40)  $ & 0.182(10) & 0.173(5) \\
\hline
\end{tabular}
\end{center}
\end{table*}

Starting with the parameters given in Table VI of \cite{dutra}
we refit $  t_{0}, t_{1}, t_{2}, t_{3}, x_{0}  $, $  x_{3}  $ and $  W  $
to the doubly-magic data and these three theoretical low-density neutron
EOS values. There is no constraint on the
neutron skin in the fit.
The $  t_{0}, t_{1}, t_{2}, t_{3}, x_{0}, x_{3}  $ and $  W  $ Skyrme parameters
are all well determined by the fit. The EOS results for the 12 EDFs are
shown in the bottom panel of Fig.~2 with the numerical results
for the parameters given in Table~I.
The six ``best-fit" results corresponding to those with a
symmetric-nuclear-matter effective mass near unity are shown in the middle panel
of Fig.~2. Inclusion of the theoretical points for low density does
not give a significant increase in the $  \chi^{2}  $ for the fit
compared to those from fits to the nuclear data alone \cite{bab}. We find
that the theoretical points for low density can be reproduced within
their error bars with all 12 of the EDFs and with similar good
results for the properties of doubly-magic nuclei. Furthermore, the
addition of the low-density points is enough to constrain the Skyrme
parameters needed for the neutron EOS. This leads to the predictions
for EOS properties and neutron skins for $^{208}$Pb and $^{48}$Ca
given in Table I.

In \cite{dutra} some of the Skyrme EDFs were eliminated because their
neutron-matter effective mass was not less than unity. But as shown
in \cite{bab} one can include the neutron-matter effective mass in the fit by
allowing the $  x_{1}  $ or $  x_{2}  $ Skyrme parameters to vary, and this
was used to set the neutron-matter effective mass to be 0.9 at a density of
0.10 fm$^{-3}$. In this Letter, we reconsider the value and
consequence of the neutron-matter effective mass based on theoretical
considerations. At low densities, the neutron-matter effective mass is close
to $  m_{n}^{*}/m \approx 1.0  $. Calculations based on chiral EFT two-
and three-nucleon interactions lead to a range $  m_{n}^{*}/m  $ = 1.0-1.1
(see Fig.~6 of \cite{heb}) and renormalization-group calculations
of the Fermi liquid parameters \cite{RGnm} give $  m_{n}^{*}/m \approx 1.0  $
up to densities $  \rho_{n} \lesssim 0.1  $ fm$^{-3}$. An effective
mass close to the bare mass or slightly increased is also expected
from the unitary regime of neutron matter at very low
densities \cite{unitary}.

To study the importance of the neutron-matter effective mass we refit
the six ``best-fit" EDFs with a constraint that the neutron matter effective
mass is unity. The results shown in the second part of Table~I and in
the top of Fig.~2.
The last line of Table~I shows the results and errors
that include a variation of 0.9 to 1.1 for the neutron-matter
effective mass. The variation in this range leads to a small change in
the neutron skins. The largest uncertainty is for the symmetry-energy
incompressibility $  K_{s}  $.

The Skyrme neutron EOS is given by the analytical expression
$$
[E/N](\rho ) =a_{n} \rho + b_{n} \rho^{\gamma} + c_{n} \rho^{2/3} + d_{n} \rho^{5/3}       
\eqno({4})
$$
where $  \gamma = 1 + \sigma  $, and $  a_{n}, b_{n}, c_{n}  $ and $  d_{n}  $ are constants
that depend on the Skyrme parameters.  The first term is from the
delta-function part that depends on $  t_{0}  $ and $  x_{0}  $, the second
term is from the density-dependent part that depends on $  t_{3}  $ and
$  x_{3}  $, the third term is the Fermi-gas kinetic energy, and the
fourth term depends on $  t_{1}, t_{2}, x_{1}  $ and $  x_{2}  $. The
neutron-matter effective mass is given by
$$
\frac{m^{*}_{n}(\rho)}{m} = \frac{c_{n}}{c_{b}+d_{n} \rho}.       \eqno({5})
$$

\begin{figure}
\includegraphics[scale=0.5]{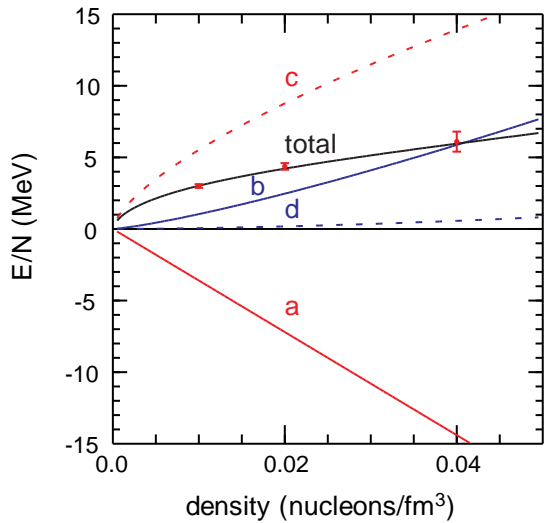}
\caption{The neutron EOS at very low density. The total result for
the Ska25 fit with a neutron-matter effective mass of 0.9 at a density of
0.10 fm$^{-3}$ is shown (black solid line)
together with the separate contributions
from the $  a_{n}  $ (red solid line), $  b_{n}  $ (blue solid line),
$  c_{n}  $ (red dashed line) and $  d_{n}  $ (black dashed line)
terms of Eq.~(4). The theoretical data
points are shown by the red points with error bars.}
\end{figure}

The very low-density results for one of the ``best-fit" cases with a
neutron-matter effective mass of unity are shown in Fig.~3.  The neutron EOS
is broken down into its four components; $  c_{n}=119  $ MeV fm$^{2}$
is fixed by the Fermi gas
model, $  d_{n}  $ is relatively small, and the two most important parameters are
$  a_{n}  $ and $  b_{n}  $. It is remarkable that the theoretical low-density EOS and
the properties of doubly-magic nuclei can all be understood with the
Skyrme ansatz for the EDF (see also \cite{hebeler13}).
The values of $  a_{n}  $ and $  b_{n}  $ depend on
$  \gamma  = 1 + \sigma   $, but the resulting neutron matter
EOS all give a good fit to the data.
There will be an uncertainty
in the neutron EOS at high
neutron density due to the dependence on
$  \sigma  $ and the effective mass within the Skyrme EDFs.
For our set of 12 EDFs and a range of
$  [m_{n}^{*}/m]  $($\rho_{n}$=0.10 fm$^{-3}$) = 0.9-1.1, we obtain
[E/N]($\rho_{n}$=0.32) = 37-51 MeV.

It has been shown that the dipole polarizability of $^{208}$Pb is
sensitive to the neutron skin of $^{208}$Pb \cite{dipole1} together
with other properties of the neutron EOS \cite{piek}, \cite{roca}. The dipole
polarizability of $^{208}$Pb was recently measured to be
$  \alpha_{D} = 20.1(6)  $ fm$^{3}$ \cite{tami}. The droplet model was used to obtain
analytical relationships between $  \alpha_{D}  $, properties of the
symmetry energy and the surface properties of
$^{208}$Pb \cite{roca}. If we use Eq.~(11) from \cite{roca} together
with the $  J  $ and $  L  $ values from the last row of Table~I, we obtain
$  \alpha_{D} = 21.3(1.3)  $ fm$^{3}$. If we use Eq.~(12) of \cite{roca}
to obtain $  R_{np}  $ from $  \alpha_{D} = 20.1(6)  $ fm$^{3}$ and
$  J = 33.2(20)  $ MeV, we obtain $  R_{np} = 0.188  $ fm with an error of 0.009
coming from Eq.~(12) in \cite{roca}, and an error of~0.011 coming from
the error in $  J  $.  Thus, the present predictions are consistent with
the measured dipole polarizability from \cite{tami}.

The $^{208}$Pb neutron skin thickness can also be obtained from the
PREX parity-violating electron-scattering experiment of
$  R_{np}=0.302\pm(0.175)_{{\rm exp}}\pm(0.026)_{{\rm model}}\pm(0.005)_{{\rm strange}}  $
fm
\cite{prex}, \cite{hor}. A PREX-II experiment has
been approved that is expected to reduce the error bar to about 0.06
fm. This and the planned parity-violating experiments on $^{48}$Ca
will be an important test of our predictions.

It will be important to see if any measured
property is inconsistent with our predictions within their error
range. If so, then a more complex form of the Skyrme functional would
be inferred. Also it will be important to carry out our analysis with
other density functional forms to see if the conclusions are robust.
The present Skyrme EDFs provide a very useful starting point for
new supernova EOS.

This work was supported in part by the NSF grant PHY-1068217, the
Helmholtz Alliance Program of the Helmholtz Association, contract
HA216/EMMI ``Extremes of Density and Temperature: Cosmic Matter in the
Laboratory", and the ERC Grant No.~307986 STRONGINT. We thank the
Institute for Nuclear Theory at the University of Washington for its
hospitality.

\end{document}